\documentclass[twocolumn,
  floatfix,
  pra,superscriptaddress,a4paper,]{revtex4-1}

\def\figwidthhalf{42mm}
\def\figwidth{85mm}
\def\Figwidth{145mm}

\usepackage[T1]{fontenc}
\usepackage[latin1]{inputenc}
\usepackage[english]{babel}

\usepackage{textcomp}
\usepackage{amsmath}
\usepackage{amsfonts}
\usepackage{amssymb}

\usepackage{units}

\usepackage{xcolor}
\usepackage{graphicx}

\usepackage{userdef_macros}

\usepackage{ifpdf}
\ifpdf
  \newcommand*{\hyperrefcolor}{violet}
  \usepackage[pdfsubject={Manuscript},pdftitle={Manuscript},pdfauthor={us},bookmarksopen=true,   
    plainpages=false,     
    colorlinks,           
    linkcolor=\hyperrefcolor,anchorcolor=\hyperrefcolor,citecolor=\hyperrefcolor,urlcolor=\hyperrefcolor,menucolor=\hyperrefcolor,filecolor=\hyperrefcolor]{hyperref}
\fi

\def\mybibstyle{apsrev4-1}

\newcommand{\theintroductorynote}{\pagenumbering{roman}
  \thispagestyle{empty}
  \section*{Introductory Note}
  This version of the manuscript contains the differences between the
  previous revision and the current one. Any change at each page is preceded
  by a unique mark, e.g. \changecntbox{1}. Changes between the revisions
  are emphasized like this: {\removedcolor\removedfont{text removed
  from the previous revision}} and {\addedcolor\addedfont{newly added text
  in the current revision}}. A list with comments explaining each change
  is appended at the end of the manuscript (see section \revisionchangesname).
  \clearpage~
  \thispagestyle{empty}
  \clearpage
}

\newcommand{\myfigureschematic}{
  \begin{figure}
    \centering
    \includecombinedgraphics{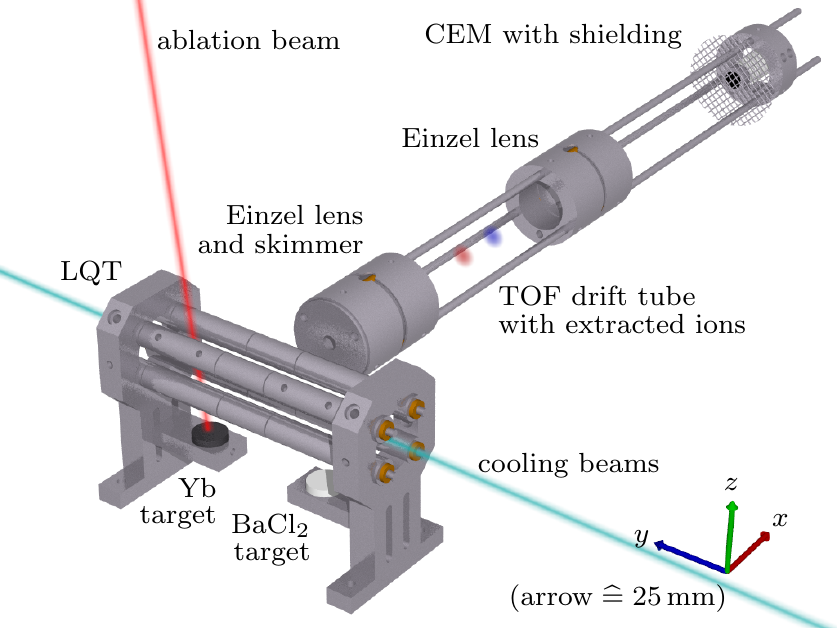}
    \caption{Schematic of the vacuum apparatus (from: \refcite{schneider:lams}).
      The linear RF trap has a field radius of
      $R_0 = \unit[6.85]{mm}$ and a length of $\unit[91]{mm}$.
      Its axis is aligned perpendicularly to the TOF drift tube to enable the
      radial extraction.
      $\atom{Yb}$ and $\atom{Ba}\atom{Cl}_2$ ablation targets are mounted
      below the RF trap.
      Laser cooling beams for $\atom{Yb}[+]$ and $\atom{Ba}[+]$ are roughly
      aligned with the axis of the RF trap and can optionally impinge under
      an angle of $\unit[45]{\degree}$ (not shown).
      The TOF drift tube contains two Einzel lenses and has a total length
      of $\unit[275]{mm}$.
      Ions are detected using a channel electron multiplier (CEM), which is
      shielded by a grounded stainless steel mesh, and the complete
      assembly is held under vacuum at a pressure of
      $\approx \unit[10^{-9}]{mbar}$.
      Optionally, a magneto-optical trap for $\atom{Ca}$ can be overlapped with
      the trapped ions (not shown).}
    \figlabel{setup}
  \end{figure}
}

\newcommand{\myfigureresultsyb}{
  \begin{figure}
    \centering
    \includecombinedgraphics{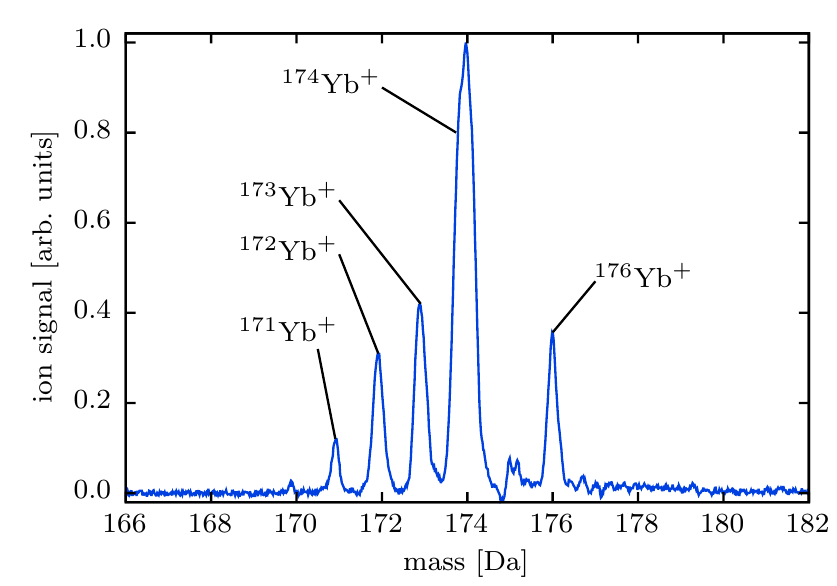}
    \caption{TOFMS spectrum for trapped $\atom{Yb}$ ablation products
      (see \refcite{schneider:lams} for more details).
      Samples consisting of $\approx 1000$ ions are loaded into the RF trap
      and $\atom[174]{Yb}[+]$ is laser cooled.
      The curve represents the average of $20$ spectra.}
    \figlabel{results:yb}
  \end{figure}
}

\newcommand{\myfiguredrive}{
  \begin{figure*}
    \centering
    \includecombinedgraphics[vecwidth=\Figwidth]{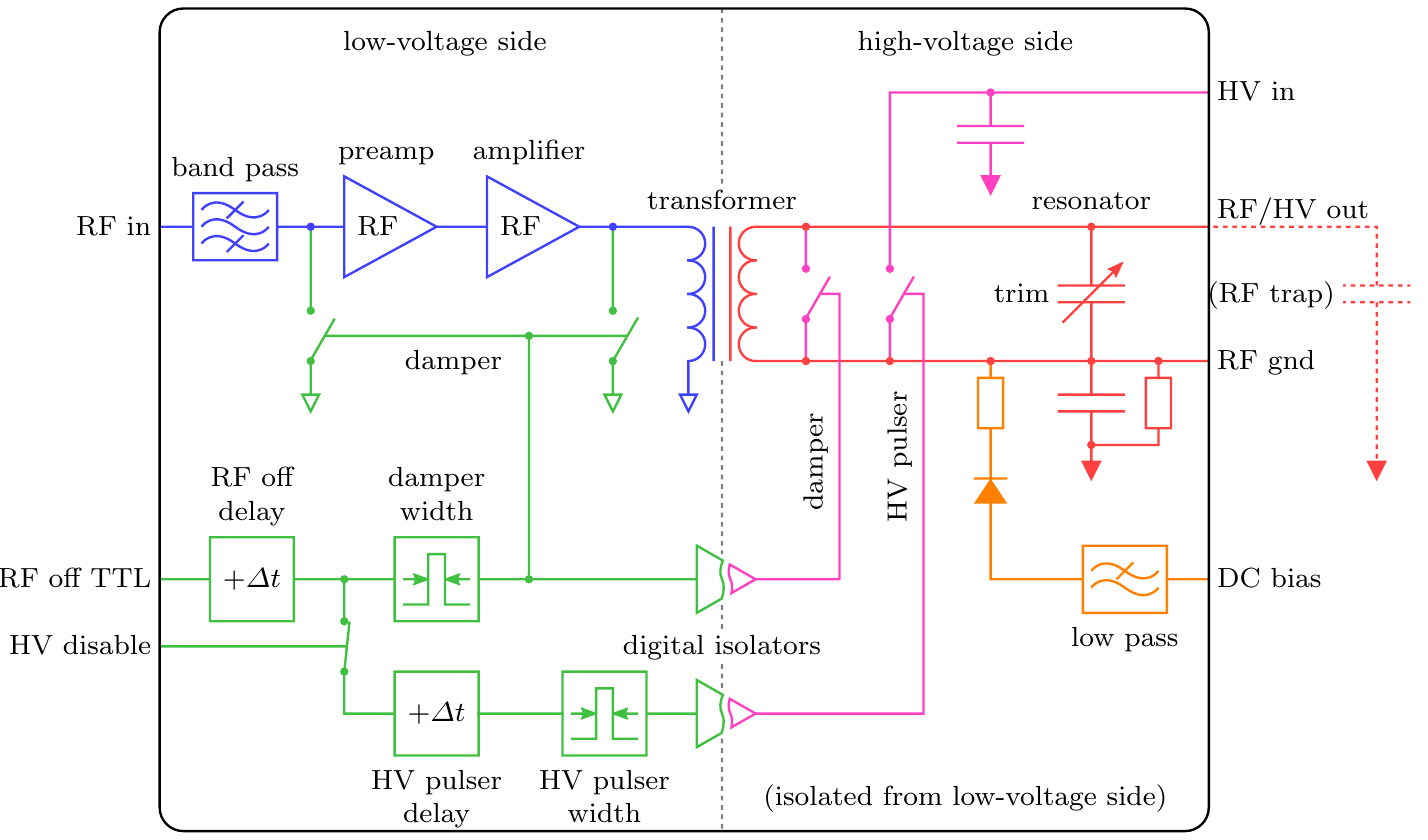}
    \caption{Block diagram of a drive unit.
      The PCB is divided into a low-voltage (left) and high-voltage
      (right) side, which are galvanically isolated from each other.
      The low-voltage side consists of RF amplifiers and primary winding
      of the RF transformer (blue) and timing and damping
      circuitry (green).
      The high-voltage side comprises the secondary winding of the transformer
      with capacitors forming the resonator (red), damping and HV pulsing
      circuitry (purple), and a $U_\text{DC}$ bias supply (orange).
      }
    \figlabel{schematics:drive}
  \end{figure*}
}

\newcommand{\myfiguredrivephoto}{
  \begin{figure}
    \centering
    \includegraphics[width=\figwidth,keepaspectratio]{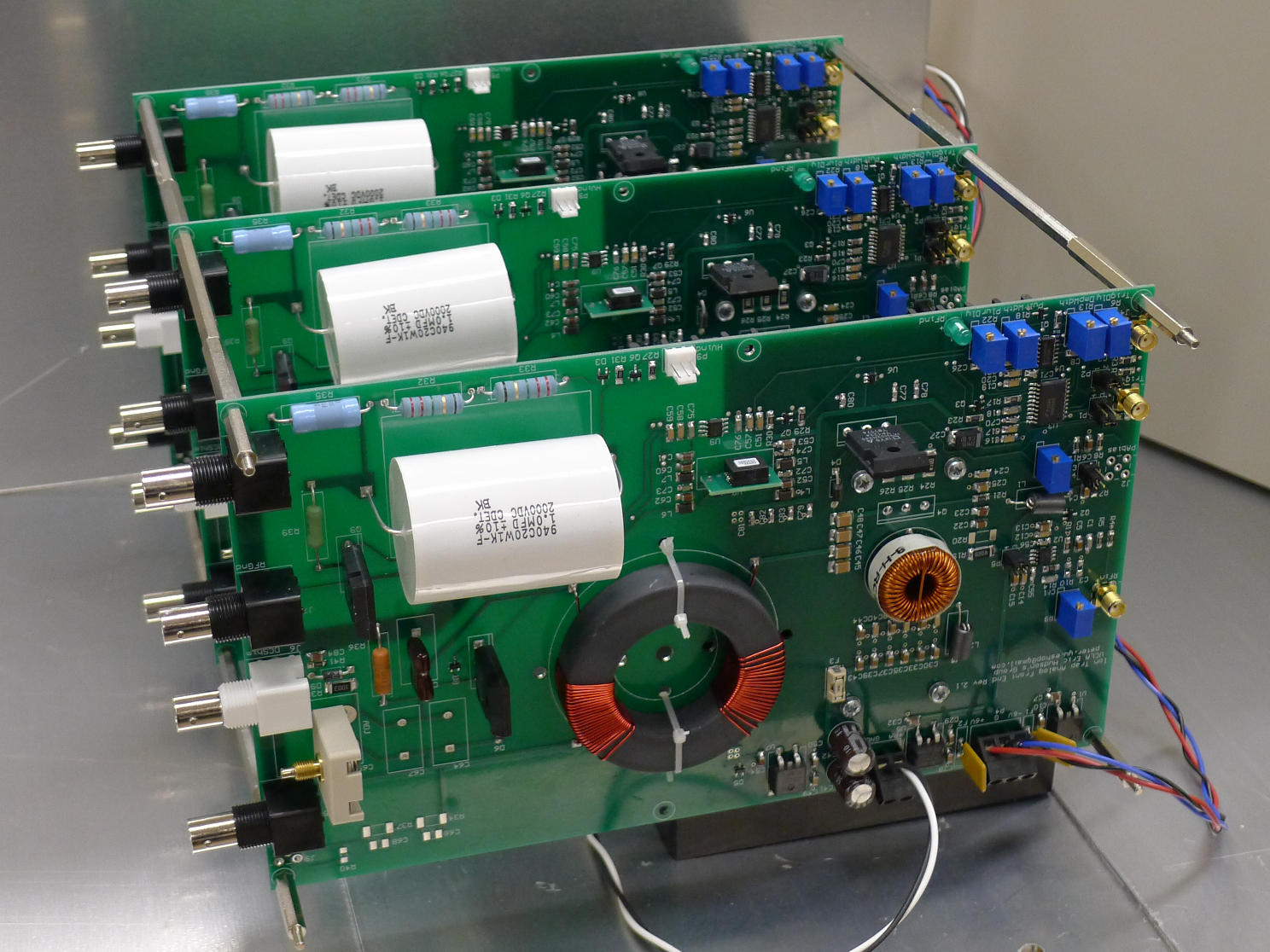}
    \caption{Photograph of three drive units assembled to fit into one
      $\unit[19]{"}$ enclosure.
      The high-voltage side can be identified on left-hand side by the light
      green PCB material (because it has no ground plane) compared to the
      dark green low-voltage side.
      The PCBs stand on the heatsinks (partially
      visible behind the front PCB) of their RF amplifier MOSFETs.
      Four copies of this assembly are required to operate the RF trap and
      TOFMS.}
    \figlabel{photo:drive}
  \end{figure}
}

\newcommand{\myfigurecontrol}{
  \begin{figure*}
    \centering
    \includecombinedgraphics[vecwidth=\Figwidth]{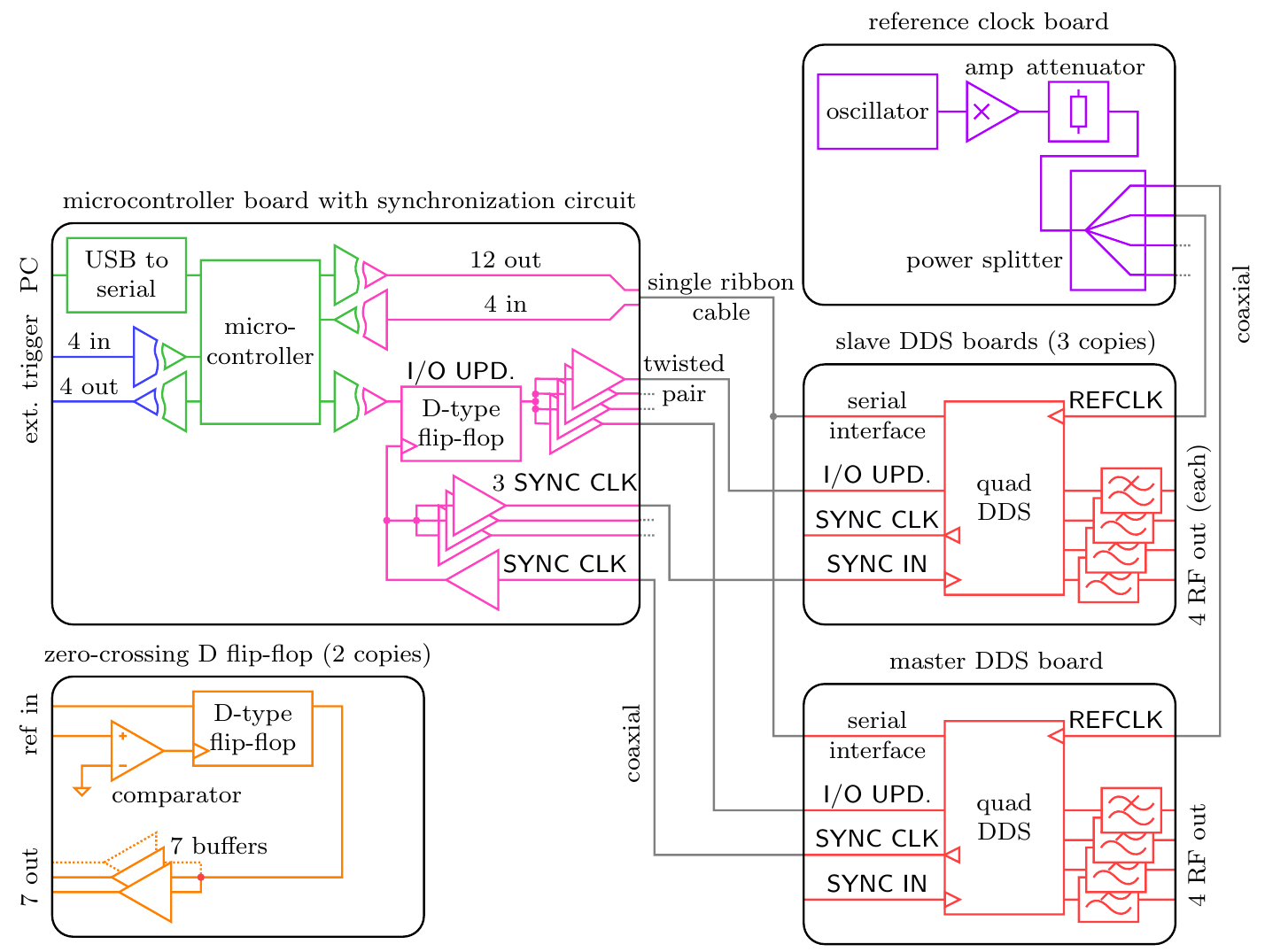}
    \caption{Block diagram of the control unit.
      The unit contains a microcontroller which is connected to a PC via USB
      (green) and possesses external trigger inputs and outputs (blue).
      The microcontroller can program four DDS devices (red) over an
      SPI-compatible serial interface.
      The DDS devices are synchronized by using the
      same reference clock (violet) and a synchronization circuit on the
      microcontroller PCB (purple).
      Two PCBs based on D-type flip-flops synchronize the TTL signal for
      initiating the HV pulses with the RF drive phase (orange).
      The microcontroller part is galvanically isolated from the rest of the
      control unit.}
    \figlabel{schematics:control}
  \end{figure*}
}

\newcommand{\myfigurecontrolphoto}{
  \begin{figure}
    \centering
    \includegraphics[width=\figwidth,keepaspectratio]{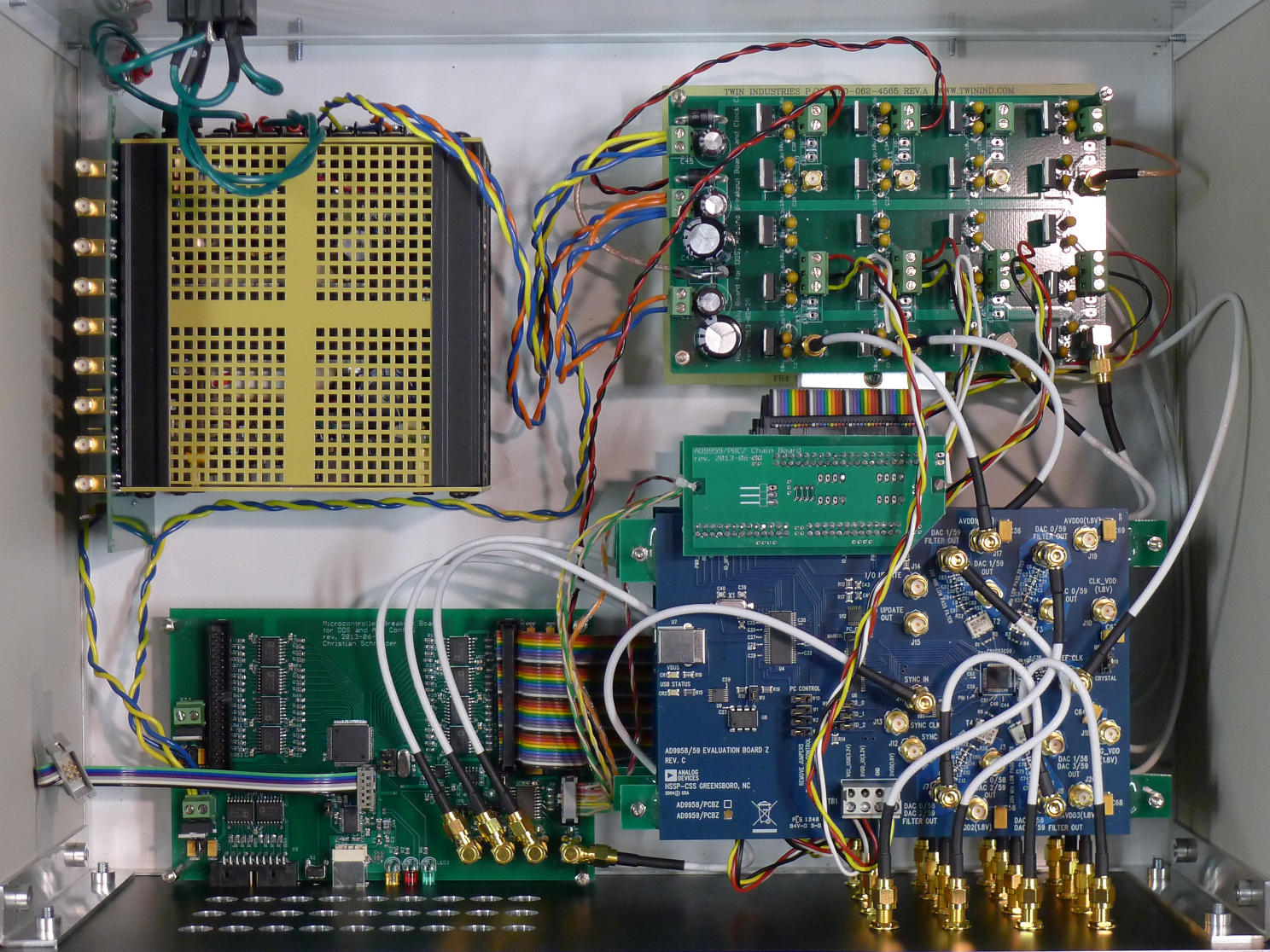}
    \caption{Photograph of the control unit.
      The microcontroller board with synchronization circuit (bottom left)
      is connected to a stack of four DDS boards (blue, bottom right).
      The serial interface is connected via the colored ribbon cable and
      small PCBs (upper left corner of DDS boards) to facilitate wiring
      and selection of $\overline{\text{\textsf{CS}}}$ and \textsf{SDO}.
      A supply (upper left) supplies a voltage regulator PCB  (upper right)
      providing power for various components.
      The reference clock board is located below the voltage regulator board
      (upper right).
      The zero-crossing D-type flip-flop boards are normally located on top
      of the microcontroller board, but have been temporarily removed.
      (One is standing upright, left of the power supply.)}
    \figlabel{photo:control}
  \end{figure}
}

\newcommand{\myfigurevoltages}{
  \begin{figure}
    \centering
    \includecombinedgraphics{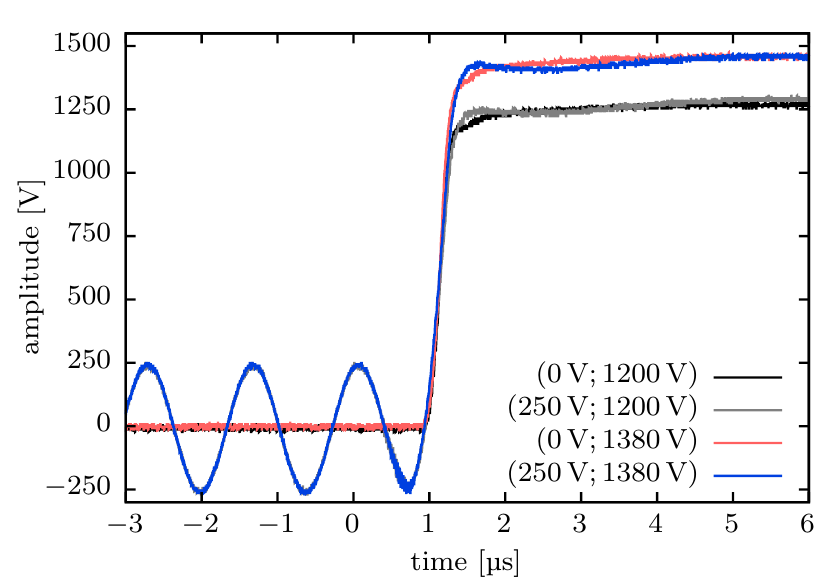}
    \caption{Typical RF voltages and HV pulses
      $(V_\text{RF}; U_\text{HV})$ at $\Omega = 2 \pi \times \unit[720]{kHz}$
      as measured at the
      inner four segments of the RF trap (see \secref{wiring}) and recorded with
      an oscilloscope.
      The time $t = 0$ refers to the the trigger of the drive units and the
      offset of the HV pulses depends on settings of the timing circuitry.}
    \figlabel{voltages}
  \end{figure}
}

\newcommand{\myfigurefeedthrough}{
  \begin{figure}
    \centering
    \subfiguremark[1,-1]{\footnotesize\colorbox{white}{a)}}{\includegraphics[width=\figwidthhalf,keepaspectratio]{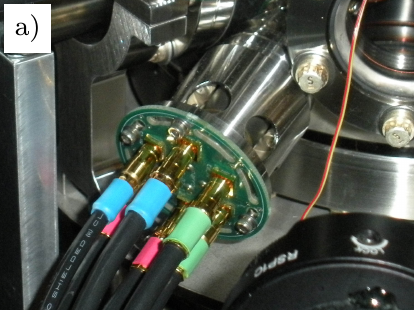}}\hfill\subfiguremark[1,-1]{\footnotesize\colorbox{white}{b)}}{\includecombinedgraphics[vecwidth=\figwidthhalf]{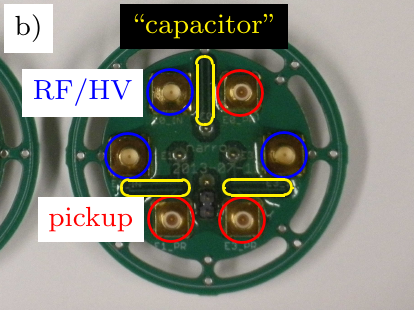}}
    \caption{(a) Feedthrough assembly for one electrode of the RF trap.
        The PCB is attached with a mechanical mount on top of a
        four-wire feedthrough and provides the interface between wires and
        SMB connectors.
      (b) Top view of the PCB.
        Three SMB connectors are used to supply the RF/HV voltages,
        while the other three SMB connectors are used to measure the input
        voltages via capacitors implemented by close-by PCB traces.}
    \figlabel{feedthrough}
  \end{figure}
}

\newcommand{\myfiguresimulationschem}{
  \begin{figure}
    \centering
    \includecombinedgraphics{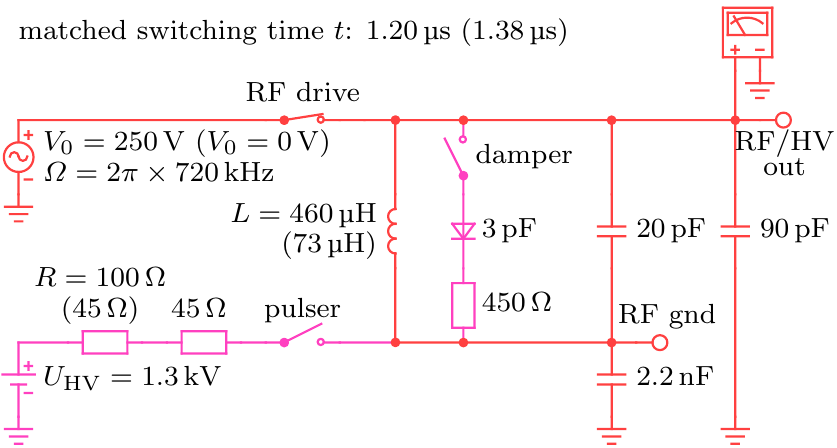}
    \caption{Simplified circuit diagram of the relevant components of a drive
      unit for simulation of the HV pulsing with Qucs \cite{Qucs2013} for an RF drive
      frequency $\Omega = \Omega_< = 2 \pi \times \unit[720]{kHz}$.
      The RF drive starts with a positive zero-crossing at $t = 0$ in the
      simulation.
      Switches change at time $t$ from open (closed) to closed (open) positions
      to disable RF drive and activate damping and HV pulsing, respectively.
      Values approximate $\Omega_<$ drive units; deviating values for $\Omega_>$
      drive units are given in parenthesis.
      Prior to matching, both drive units use $R = \unit[100]{\Ohm}$ and
      $t \approx \unit[1.389]{\micro s}$.
      Coarse matching leads to $R = \unit[45]{\Ohm}$ for the
      $\Omega_>$ drive units and the switching times given in the diagram.}
    \figlabel{simulation:schem}
  \end{figure}
}

\newcommand{\myfiguresimulation}{
  \begin{figure}
    \centering
    \includecombinedgraphics{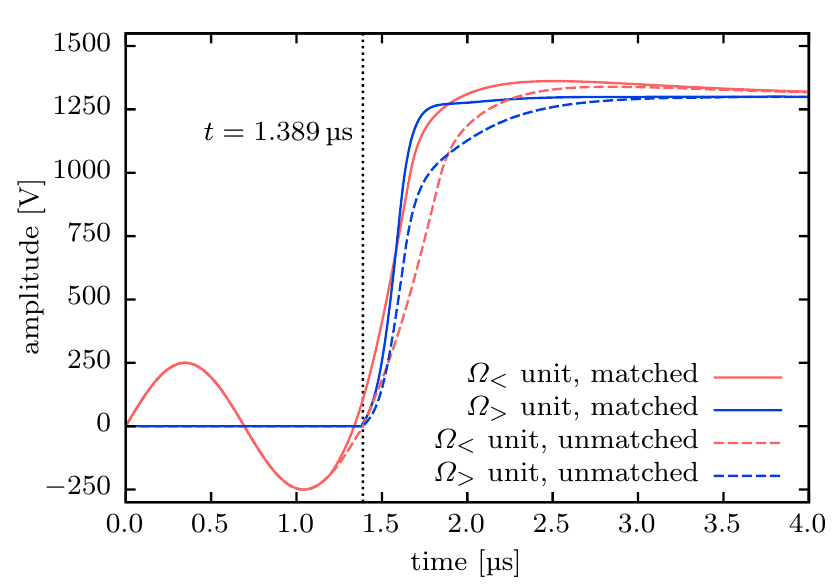}
    \caption{Simulated output voltages of the $\Omega_<$ and $\Omega_>$
      drive units for the coarsely matched case (solid curves) compared to
      the values prior to matching (dashed curves); compare \figref{simulation:schem}.}
    \figlabel{simulation}
  \end{figure}
}

\begin{document}
\newcommand*\includecombinedgraphics[2][]{\includegraphics{#2}}
\renewcommand*\subfiguremark[3][]{#3}

\pagenumbering{arabic}

\title{Ion trap with integrated time-of-flight mass spectrometer}

\author{Christian Schneider}
\email{christian.schneider@physics.ucla.edu}
\author{Steven J. Schowalter}
\author{Peter Yu}
\author{Eric R. Hudson}
\affiliation{Department of Physics and Astronomy,
             University of California,
             Los Angeles,
             California 90095, USA}

\newcommand{\mytextabstract}{
  Recently, we reported an ion trap experiment with an integrated
  time-of-flight mass spectrometer (TOFMS) \cite{schneider:lams} focussing on the
  improvement of mass resolution and detection limit due to sample preparation
  at millikelvin temperatures.
  The system utilizes a radio-frequency (RF) ion trap with asymmetric drive for
  storing and manipulating laser-cooled ions and features radial extraction
  into a compact $\unit[275]{mm}$ long TOF drift tube.
  The mass resolution exceeds $m / \Delta m = 500$, which provides isotopic
  resolution over the whole mass range of interest in current experiments and
  constitutes an improvement of almost an order of magnitude over other
  implementations.
  In this manuscript, we discuss the experimental implementation in detail,
  which is comprised of newly developed drive electronics for generating the required
  voltages to operate RF trap and TOFMS, as well as control electronics for
  regulating RF outputs and synchronizing the TOFMS extraction.
}

  \begin{abstract}
    \mytextabstract
  \end{abstract}

  \maketitle

\section{Introduction}

Experiments with molecular ions in radio-frequency (RF) ion traps have
rapidly evolved in physics and chemistry in recent years.
Such experiments focus on the production of molecules \cite{tong:rot:vib:ion};
their cooling \cite{hudson:cold:molecules,schneider:rovib:cool,
staanum:rot:cool:mgh,hansen:rot:cool,lien:rot:cool,rellergert:vib:cool};
reactions of trapped ions with (untrapped) neutral reactants
\cite{chang:conformers,DePalatis2013,willitsch:collisions,Deb2015};
spectroscopy of molecular ions
\cite{chen:spec:bacl,schowalter:tof,Seck2014,Puri2014,dunning:in:prep}; or
precision measurements \cite{Ni2014}.
As opposed to neutral molecules, molecular ions allow for easy trapping and,
optionally, sympathetic cooling of their translational degrees of
freedom with co-trapped, laser-cooled atomic ions
\cite{larson:symp:cool,baba:symp:cool,Rugango2015}.
Since ion trapping is largely species independent, it is important to have a
robust means to identify the trapped ions.
This identification can be used to deduce reaction properties, such as reaction
rates or branching ratios, or in destructive spectroscopy techniques involving
e.g. photo-dissociation.

Closely related are experimental efforts with hybrid atom--ion traps
\cite{Smith2005,grier:collisions,
haerter:atom:ion:review,Sias2014,willitsch:atom:ion:review}.
Their implementation typically involves an RF ion
trap for confining ions, which are overlapped with a
cold cloud of atoms \cite{Smith2005,grier:collisions} or a
Bose-Einstein condensate \cite{zipkes:ion:bec,schmid:ion:bec}.
All-optical hybrid traps for atoms and ions are also in development,
in which the RF trap can be turned off during certain experimental sequences
\cite{schneider:dipole:trap,schneider:dipole:trap:ext,
enderlein:optical:lattice,Huber2014}.
A chief application of these systems is the study of cold/ultra-cold collisions
and
reactions of ions and atoms \cite{zipkes:collisions,sullivan:reaction:ca2,
rellergert:reaction:ybca,Hall2011,Ratschbacher2012,sullivan:reaction:baca,
Hall2012,Sivarajah2012}.
Conclusions on the reaction mechanisms can be drawn by trapping and, again,
subsequently analysing charged reaction products.

Mass spectrometry (MS) constitutes a powerful and straight-forward way
for the analyses of product ion samples in such experiments.
While mass spectrometers with resolutions exceeding $m / \Delta m = 100,000$
are commercially available \cite{makarov:orbitrap}, such systems are usually bulky,
expensive and difficult to integrate in experiments with cold ions and atoms.
Hence, various techniques have been used to allow for discriminating
different ions species in such experiments.
These include mass filtering
\cite{paul:massfilter}; resonant excitation of the
secular ion motion \cite{baba:symp:cool,drewsen:tickling,ostendorf:symp:cool};
laser-induced fluorescence techniques \cite{schiller:mol:sim,zhang:mol:sim,
staanum:rot:cool:mgh,kahra:mgh,chang:conformers}; or
integrated time-of-flight mass spectrometers (TOFMSs) \cite{schowalter:tof,
Ni2014,Seck2014,schneider:lams,Deb2015}.
The former techniques are applicable without changes to the vacuum systems,
but complicated due to nonlinear resonances of the RF trap
\cite{alheit:resonances}, complex interpretation of the resulting spectra,
demanding requirements on laser cooling including formation of ion Coulomb
crystals, and/or the required molecular dynamics simulations.
This is in particular true, if the sample contains a variety of
potentially unknown ion species.
As an alternative, integrated TOFMSs have proven to be unambiguos, be conceptually
simple, be widely applicable, and provide a relatively high, sometimes isotopic, mass
resolution.

An early implementation of an RF trap with integrated TOFMS is given
in \refcite{schowalter:tof} and reaches mass resolutions of typically
$m / \Delta m \sim 50$.
The linear RF trap is operated symmetrically (with RF voltages of opposite sign
at neighboring electrodes) using two center-tapped RF transformers.
Pulsed application of slightly different high voltages (HVs) to the center taps
of the transformers creates the two-stage electric field of a Wiley-McLaren
TOFMS \cite{wiley:tof} and extracts ions radially
\cite{jolliffe:tof:radial,franzen:tof:radial} into a TOF drift tube.
The radial extraction of this implementation has the advantage of a more
compressed sample compared to an extraction along the axis of the RF
trap, but sacrifices some of the gain in mass resolution due to the presence of
RF ringing during the extraction.
This system has been used to perform photo-dissociation spectroscopy
of $\atom{Ba}\atom{Cl}[+]$ \cite{schowalter:tof}, $\atom{Sr}\atom{Cl}[+]$
\cite{Puri2014}, and $\atom{Dy}\atom{Cl}[+]$ \cite{dunning:in:prep}.

A later TOFMS implementation using a six-rod quadrupole RF trap
with radial extraction is described in \refcite{Ni2014} and has been used
for spectroscopy of $\atom{Hf}\atom{F}[+]$.
The RF trap is driven at an exceptionally low drive frequency of about
$\unit[50]{kHz}$, such that drive voltages are generated without resonant
enhancement which facilitates the application of the TOF extraction voltages.
Although the six-rod RF trap has a different motivation in
this system, such a trap geometry has the potential advantage of
being able to use separate electrodes for RF and HV:
while two opposing rods can be driven with RF, HVs can be applied
to the other four rods and ions are extracted through a gap between two
neighboring HV electrodes.
A drawback for certain applications would be the reduced optical
access for establishing, for example, an atom trap.

Another TOFMS implementation with axial extraction has been used to perform
spectroscopy of $\atom{Al}\atom{H}[+]$ \cite{Seck2014}.
While the mass resolution is limited, the implementation is appealing because of
its simplicity and the separation of RF and HV electrodes.

Recently, a TOFMS with radial extraction has been demonstrated
\cite{Deb2015}, in which the RF trap is operated with digital HV pulses
instead of sinusoidal voltages \cite{Kjargaard2001,Kjargaard2002}.
This concept appears compelling, because the abscence of RF voltages prevents
the associated ringing and digital HV pulses can be directly reused for
extraction into the TOFMS, however, the mass resolution in the given
implementation remains $m / \Delta m \le 90$.
Further, the digital drive leads to enhanced micromotion and increased
requirements on phase matching between different electrodes, which is
disadvantageous during experimental cycles requiring cold ion samples.

In 2014, we demonstrated an RF trap with integrated TOFMS \cite{schneider:lams},
which is loosely based on \refscite{makarov:rf:supply}{schowalter:tof}.
The TOFMS has a significantly higher mass resolution
($m / \Delta m > 500$) than other implementations.
The RF trap is operated asymmetrically and can be actively damped during
the application of the HV extraction pulses to prevent ringing.
The basic setup and the effect of laser cooling on mass spectrometry has been
sketched in \refcite{schneider:lams}; here, we focus on the technical details.

In the following, we first explain the electronics for the RF trap/TOFMS in
\secref{implementation}.
This section is subdivided into an overview of the experimental apparatus
(\secref{overview}), the description of drive units generating the RF
voltages and HVs (\secref{drive}) and a control unit (\secref{control}), and
remarks on the wiring of the RF trap and measuring of the voltages
(\secref{wiring}).
Subsequently, in \secref{discussion}, we conclude with a brief discussion of the
performance of the system (\secref{performance}) and potential future
improvements (\secref{improvements}).

\section{Experimental Implementation}
\seclabel{implementation}

\subsection{Overview}
\seclabel{overview}

\iftwocolumn{\myfigureschematic}{}

The apparatus consists of a segmented linear RF trap, and a
basic, Wiley-McLaren type TOFMS (see \figref{setup}).
More details are given in \refcite{schneider:lams}.
Briefly, the RF trap is driven asymmetrically with one pair of diagonally opposing
electrodes at RF voltage (amplitude $V_\text{RF} = V_0$) and the other pair at
RF ground ($V_\text{RF} = 0$).
We can choose between a drive frequency $\Omega$ of either
$\Omega_< \approx 2 \pi \times \unit[720]{kHz}$ or
$\Omega_> \approx 2 \pi \times \unit[1.8]{MHz}$ depending on the pair
of electrodes being driven and reach RF amplitudes of up to
$V_0 \approx \unit[750]{V}$.

For the extraction into the TOF drift tube, the RF voltages are turned off.
Subsequently, the electrodes are pulsed to DC HVs $U_\text{HV}$
with a $\unit[10]{\%}\rangeto\unit[90]{\%}$ rise time of $\approx \unit[250]{ns}$.
The HV is applied such that a two-stage electric field is
established \cite{wiley:tof}, which radially extracts the cold atoms and
molecules from the RF trap into the TOFMS
\cite{jolliffe:tof:radial,franzen:tof:radial}.
This is accomplished by applying a slightly lower HV to the electrodes which are
closer to the TOFMS ($U_\text{HV} = U_1 = \unit[1.2]{kV}$) than to the ones
that are farther ($U_\text{HV} = U_2 = \unit[1.4]{kV}$).

This scheme leads to a noticable complication:
While the same RF voltage $V_\text{RF}$ is required on diagonally
opposing electrodes for trapping, different $U_\text{HV}$
must be applied to these electrodes for extraction into the TOFMS.
In total, four different configurations of voltages $(V_\text{RF}; U_\text{HV})$
must be generated: $(0; U_1)$, $(0; U_2)$, $(V_0; U_1)$, and
$(V_0; U_2)$.
Additionally, different low-voltage DC voltages $U_\text{DC}$ need to be
superimposed with $V_\text{RF}$ on some segments
to compensate micromotion and provide axial confinement.

The developed circuit, referred to as the drive unit (see \secref{drive}) in what follows,
generates one triplet of voltages $(V_\text{RF}; U_\text{HV}; U_\text{DC})$.
We typically drive each segment of the RF trap with its own
drive unit such that a total of twelve copies of these drive units is required.
This choice has the benefit of a large degree of freedom in the applied
voltages, but comes at the price of a relatively high complexity for matching
both the RF voltages (frequency,
phase and amplitude) and HV pulses (timing, slope and amplitude).
A control unit (see \secref{control}) outputting up to 16 synchronized RF signals
with adjustable frequency, phase and amplitude and incorporating
synchronization circuits plays a key role in this matching.
In principle, the number of drive units could be reduced to four, if a
non-segmented RF trap with external DC electrodes was used.

\subsection{Drive Unit}
\seclabel{drive}

\iftwocolumn{\myfiguredrive}{}

\iftwocolumn{\myfiguredrivephoto}{}

\iftwocolumn{\myfiguresimulationschem}{}

\iftwocolumn{\myfiguresimulation}{}

\iftwocolumn{\myfigurecontrol}{}

\iftwocolumn{\myfigurecontrolphoto}{}

The block diagram of the drive unit is given in
\figref{schematics:drive} and a photograph is shown in \figref{photo:drive}.
The entire unit is implemented on a single printed-circuit board (PCB).
The incoming RF signal passes an RC-bandpass filter and is amplified by a
preamplifier (Analog Devices
ADA4897-1) and
power amplifier (MOSFET: Microsemi ARF460AG).
The power amplifier is operated in class A mode to minimize distortion and
drives the primary side of a toroidal transformer (ferrite:
material 61, $\mu = 125$, $\unit[1.4]{"}$ outer diameter).
An impedance of only a few Ohms is chosen to limit the supply voltage
of the amplifier to $\unit[24]{V}$ and the number of turns of the transformer
to a few ten.

The secondary side of the transformer together with a trim capacitor, PCB
traces, cables, the vacuum feedthrough, and the RF trap segment
form a $LC$-resonator ($L \sim \unit[(0.5\rangeto1.0)]{mH}$ depending
on the desired $\Omega$ and $C \sim \unit[100]{pF}$).
The trim capacitor ($\unit[(12\rangeto 48)]{pF}$) allows for fine-tuning the
resonance frequency and matching the frequencies of the respective drive
units.
While one end of the secondary winding of the transformer is connected
to the segment of the RF trap, the other end is RF grounded via a capacitor
($\sim 20 \times C$) which allows biasing of the RF trap segment with both
a positive $U_\text{DC}$ and $U_\text{HV}$.

$U_\text{DC}$ is applied via a low-pass filter and an HV diode (as a protection
from $U_\text{HV}$) to RF ground.
Similarly, a $U_\text{HV}$ pulse is applied to RF ground by activating a MOSFET
(IXYS IXTH1N250) which is supplied by a HV power supply and
bypassed with a capacitor ($\unit[1]{\micro F}$).
As a result of this design, not only the RF trap segment but the entire secondary
side of the drive unit is biased with $U_\text{DC}$ permanently and
$U_\text{HV}$ during pulsing.
While the RF transformer naturally isolates the amplifier, any digital signals
must be passed by digital isolators (Analog Devices ADUM series) to the
secondary side and are supplied by a DC--DC converter.
As the DC--DC converter introduces noise on the RF drive
around its switching frequency plus harmonics, a battery can optionally buffer
the voltage and allows to switch off the DC--DC converter during experiments.

Turning off the input RF signal to the resonant circuit would result in a
ring-down of the RF voltage over a few $\unit[10]{\micro s}$.
Analogously, the application of the HV pulse would cause significant ringing
on top of the HV pulses for extraction into the TOFMS.
To remedy these effects, we implemented an active damping of the
resonator on both its primary and secondary side using MOSFETs
(IXYS IXTH02N250), which effectively shorts the windings of the RF transformer.
A single input TTL signal initiates both the damping and the HV pulse.
Logic circuitry allows adjustment of delay and duration of the
damping and the HV pulse individually and is used to match the timings across all
drive units.

Still, activating the damping and HV pulsing sequence at random times and, hence,
random phases of the RF drive would result in ringing and varying slopes
of the HV pulses.
 Hence, the control unit (see \secref{control}) contains circuitry to
generate TTL signals which are synchronized with the RF drive phase to switch
roughly at the zero-crossing of the RF drive.

Further, drive units belonging to $\Omega_<$ and $\Omega_>$ exhibit initially
different HV pulsing characteristics caused by the different
inductances, $L$, of their RF transformers.
In order to assist matching, we perform a simulation with
the circuit simulator Qucs \cite{Qucs2013}.
As an example, the circuit diagram for the case $\Omega = \Omega_<
= 2 \pi \times \unit[720]{kHz}$ is presented in \figref{simulation:schem}.
Without matching, (1) both drive units use a resistor $R = \unit[100]{\Ohm}$ to limit the
HV pulsing current and (2) the HV pulses of both the $\Omega_<$ and $\Omega_>$
are activated at the positive zero-crossing of the RF drive (at time
$t \approx \unit[1.389]{\micro s}$ in the simulation).
The corresponding HV pulse forms are shown in \figref{simulation} (dashed
curves) and are in good agreement with the experimentally observed pulse forms.

Coarse matching can be achieved by two modifications:
(1) Using $R = \unit[45]{\Ohm}$ for the $\Omega_>$ drive units increases the
HV pulsing current and removes the ``kink'' in the pulse form at
$\sim \nicefrac{3}{4} U_\text{HV}$, including the shallower slope.
(2) Activating the pulses of the $\Omega_<$ drive units sightly before the
positive zero-crossing of the RF drive at $t = \unit[1.20]{\micro s}$
leads to a steeper slope and approximates the HV pulses of the $\Omega_>$ drive
units better.
(To match the timing, the HV pulses of the latter drive units are also activated
slighly earlier at $t = \unit[1.38]{\micro s}$.)
The coarsely matched HV pulse forms are also shown in \figref{simulation}
(solid curves).
The experimental system performance is discussed later (see \secref{performance}
and \figref{voltages}).

Matching using modification (2) removes the possibility of
having a delay between turning off the RF drive and activating the pulses and,
hence, ``time-lag energy focussing'' as described for the original Wiley-McLaren
TOFMS \cite{wiley:tof} is not possible.
However, this is permissible without sacrificing TOFMS resultion, because
the ions will be sufficiently cold (usually laser cooled) in the intended
experiments.

Drive unit PCBs (for the three segments of an electrode; see
\figref{photo:drive}) are
housed by threes in $\unit[19]{"}$ rack mount enclosures, which shield the
sensitive RF amplifier circuits.
Further, the design of the enclosures involves thermal management to keep the components,
particularly those of the resonator, close to ambient temperature and minimize
thermal drifts.
As a result, the drive units have a sufficient passive stability such that no
repeated matching is required.

\subsection{Control Unit}
\seclabel{control}

The control unit has three main purposes:
(1) it provides synchronized RF signals for the drive units; (2)
it allows for computer control of the RF parameters (frequency, phase,
amplitude) and certain experimental sequences (extraction into the TOF,
loading of the RF trap via laser ablation); and (3) synchronization of the
RF damping and HV pulses with the RF drive phase.

A block diagram of the control unit is depicted in \figref{schematics:control}
and a photograph in \figref{photo:control}.
The unit uses a microcontroller (Atmel ATmega 2560), which is connected over
a serial-to-USB interface (FTDI FT232R) to a computer.
The microcontroller is powered over USB and its digital inputs/outputs
are galvanically isolated by digital isolators (Analog Devices ADUM series),
which provides both reduction of noise and protection of the computer in
case of a malfunction.

Four external digital inputs and four outputs allow interfacing with
external TTL-compatible devices.
All four inputs are assigned interrupts such that externally triggered
events meet hard real-time criteria.
As an example, one of the TTL inputs can be used to turn off the DDS channels.

The RF signals are generated by four direct digital synthesis (DDS) devices
(Analog Devices AD9959) each having four channels.
The DDS devices allow for a sample rate of $\unit[500]{MSPS}$ and
individual frequency, phase, and amplitude control of each channel.
For simplicity, we utilize evaluation boards (Analog Devices AD9959/PCB).

Synchronization of the DDS devices is subject to three conditions.
First, all four devices must receive the same reference clock
(\textsf{REFCLK}) signal.
This is achieved with a $\unit[500]{MHz}$ reference oscillator (Crystek
RFPRO33-500.000), followed by an amplifier (Mini-Circuits ZX60-33LN),
attenuators (Mini-Circuits VAT series), a power splitter (Mini-Circuits
ZB4PD1-500), and coaxial cables of equal length.

Second, the synchronization clocks (\textsf{SYNC CLK} = \textsf{REFCLK}/4 in
our case) of the digital interface of the DDS devices must be
synchronized across all four DDS devices.
Without synchronization, the \textsf{SYNC CLK}s of the four DDS devices will
have mutual phase differences of a random integer-multiple of
$\unit[90]{\degree}$ upon initialization.
We chose automatic mode synchronization and supply the buffered
(Texas Instruments 74LVC541)
\textsf{SYNC CLK} output of the master device to the \textsf{SYNC IN}
of the slaves.
This results in a deterministic phase difference of an integer-multiple
of $\unit[90]{\degree}$ between master and slave devices upon initialization,
which can be removed by programming certain registers of the slave
DDS devices.

Third, the \textsf{I/O UPDATE} signal upon which data in the serial I/O buffer
of the DDS devices are transferred into active registers must be
synchronized with the \textsf{SYNC CLK}.
The \textsf{I/O UPDATE} originates from the microcontroller and is
initially unsynchronized, as the microcontroller has its own $\unit[16]{MHz}$
clock.
An edge-triggered D-type flip-flop (Texas Instruments SN74LVC1G79) passes the
unsynchronized
signal upon the rising-edge of the master's \textsf{SYNC CLK} to the DDS
devices and hence ensures an appropriately synchronized \textsf{I/O UPDATE}
signal.
(This is the reason for use of the master's \textsf{SYNC CLK} instead
of \textsf{SYNC OUT} for synchronization, because the latter requires
activation over the serial interface upon reset.
This, however, would not be possible, because \textsf{I/O UPDATE} would not
work without the clock for the flip-flop.)

The digital interface between the microcontroller and DDS devices is serial
peripheral interface (SPI) compatible.
Each DDS device uses an individual chip select ($\overline{\text{\textsf{CS}}}$)
and serial data out (\textsf{SDO}) pin of the microcontroller; any other
signal is connected in parallel across all DDS devices.
This layout manages without active components (e.g. multiplexers) and limits
the required number of wires.
Further, the DDS devices feature a profile selection over four designated pins,
which allows to switch between certain parameters within one clock cycle without
reprogramming registers over the SPI interface.
The corresponding pins are also wired in parallel across all DDS
devices and, currently, enable us to turn off all DDS channels simultaneously.

The last component of the control unit is a zero-crossing D-type flip-flop
PCB for synchronizing the damping and HV pulse with the RF drive
phase.
Similar to the \textsf{I/O UPDATE} case, an edge-triggered D-type flip-flop
(Texas Instruments SN74LVC1G79) passes an externally provided, unsynchronized
trigger signal to the drive units upon the zero-crossing of a reference RF signal
from a spare DDS channel using a comparator (Analog Devices ADCMP601).
This translates the excellent phase control of the DDS channels to a control
over the timing of the damping and HV pulsing.
In total, we use two such zero-crossing D-type flip-flop PCBs with their own
DDS reference channels to control the timining of the $\Omega_<$ and $\Omega_>$
drive units individually.

\subsection{Wiring and Measuring}
\seclabel{wiring}

\iftwocolumn{\myfigurefeedthrough}{}

The outputs of the drive units are connected to the vacuum chamber over
$\approx \unit[175]{cm}$ long low-capacitance coaxial cables (RG-62,
$\unit[42]{pF/m}$), which represent the largest single contribution to the total
capacitance of the resonator.
These cables allow for sufficient spatial separation between vacuum chamber and
optical setup such that thermal issues are prevented.

The cables have $\unit[75]{\Ohm}$ mini-SMB connectors and are plugged into
one of four assemblies on the vacuum chamber (see \figref{feedthrough}).
Each assembly consists of a PCB with mechanical mount, which is attached to a
standard $\unit[1.33]{"}$-CF four-wire vacuum feedthrough.
The PCBs interface the wires of the feedthroughs with the help of receptacles
(Mill-Max 0492-0-15-15-13-14-04-0) to SMB connectors and
provide the ground connection to the vacuum chamber.
On the vacuum side, three wires per feedthrough are connected to the three
segments of an electrode (leaving one wire per feedthrough unused).

The interface PCBs also include capacitive pickup traces close to RF/HV traces,
which sample a small fraction of the RF/HV voltage supplied to the segments.
The pickup signal allows the measurement of the input voltages at each segment
with a
probe ratio of $\approx 1000 : 1$ (as measured with an oscilloscope with
$\unit[1]{M\Ohm}$ impedance and typical cable lengths).
The ratios are calibrated against an HV probe (Agilent 10076B)
to better than $\unit[1]{\%}$ (relative).
In order to prevent arcing, in particular across the pickup traces, the PCBs
are sealed with conventional two-component epoxy adhesive and
tested with proof voltages exceeding $\unit[3]{kV}$.

\section{Discussion}
\seclabel{discussion}

\subsection{System Performance}
\seclabel{performance}

\iftwocolumn{\myfigurevoltages}{}

Typical output voltages as measured on the inner four
segments of the RF trap are given in \figref{voltages}.
The RF amplitudes can be matched using such pickup signals to at least the
$\unit[1]{\%}$ level and their mutual phase offsets can be minimized below
$\approx \unit[0.1]{\%}$.
The HV slopes can be matched to time differences $<\unit[5]{ns}$ and show
jitters $<\unit[5]{ns}$.
Ultimately, RF phase offsets or amplitudes are fine-tuned using feedback on
trapped ion Coulomb crystals; individual $U_\text{HV}$ values are optimized for
highest TOFMS detection yields and mass resolution.

\iftwocolumn{\myfigureresultsyb}{}

The detailled analysis of the performance of the TOFMS is described elsewhere
\cite{schneider:lams}.
An example TOF mass spectrum after ablating the $\atom{Yb}$ target and
laser cooling one of the $\atom{Yb}$ isotopes is depicted in \figref{results:yb}
(see \refcite{schneider:lams} for details).
The spectrum shows resolved peaks for all natural isotopes of $\atom{Yb}$.
The peak heights do not represent the natural abundances of the isotopes,
because the sample preparation is biased towards the laser-cooled isotope.
Additionally, the laser-cooled ions can more efficiently sympathetically cool
heavier ions compared to lighter ones.
The small peak around $\unit[175]{Da}$ is likely an electronic artifact
and/or due to $\atom{Yb}\atom{H}[+]$, as there is no naturally occuring
$\atom[175]{Yb}$.
Still, provided the ion detector is not saturated, the peak heights represent
the abundance of the different isotopes in the sample.

\subsection{Future Improvements}
\seclabel{improvements}

The design choice of a segmented linear RF trap with twelve individual drive
units complicates the matching of RF drive voltages and HV pulses.
Further, the freedom in choice of RF phase and amplitude for each segment is
restricted, because the mutual capacitance between the segments leads to strong
coupling of the three resonators.
Future versions will use a non-segmented linear RF trap with continuous RF
electrodes and external DC endcap electrodes, which reduces the number
of drive units to four.

Further, the CEM ion detector has a comparatively small
aperture and replacing it with a multi-channel plate (MCP) promises an improved
detection limit (close to unity) and, especially, improved linearity.

Currently, the slope of the HV pulses is intentially limited to a
$\unit[10]{\%}\rangeto\unit[90]{\%}$ rise time of $\approx \unit[250]{ns}$
by the resistor $R$.
This reduces the stress on the switching components and promises a longer
lifetime, however, it might also limit the mass resolution.
A (slight) improvement of mass resolution might be achieved by using
larger HV pulse currents without the resistor.

Ions in the RF trap have typical secular frequencies of $\omega = 2 \pi \times
\unit[(5 \rangeto 100)]{kHz}$.
The ions' secular motion in the RF trap can be heated by various mechanism
\cite{wineland:bible}, for example, by low-frequency
electronic noise around $\omega$.
This noise is reduced by the abovementioned filtering and battery buffering.
Additionally, noise on the RF drive in a region $\Omega \pm \omega$ can heat
the ions' secular motion, which is particularly important to consider when using
digital RF sources such as a DDS device.
The frequency spectrum of the drive units shows an average noise floor at the
$\le \unit[-90]{dBc}$ level at $\approx \Omega \pm \omega$, however,
due to the digital character of the DDS devices, peaks with amplitudes of
typically $\le \unit[-70]{dBc}$ are present.
The location and amplitude of these peaks depends strongly on the generated
output RF frequency.
While current experiments have not been limited by heating of the RF trap, the
noise might become a concern in future experiments.
As a large number of indepently controllable RF sources is required for the
setup described in this manuscript, it appears unrealistic to use multiple
high-quality RF generators typically used with RF traps.
As a cheaper solution, analog oscillators could be phase-locked
to the outputs of the DDS devices and provide low-noise RF sources without
losing the phase and frequency control provided by the DDS devices.

In principle, the demonstrated RF trap with integrated TOFMS is suited for work
on heavier molecules.
Singly-charged molecules of masses $m \sim \unit[1000]{Da}$ could be
trapped, sympathetically cooled \cite{schiller:mol:sim}, and resolved by the
TOFMS, which opens up experiments on a variety of volatile organic compounds,
the amino acids, explosive agents, some peptides,
and heavier biomolecules such as nucleic acids.
The mass range could be significantly extended by permitting multiply-charged
molecules as in \refcite{offenberg:proteins}.
For such experiments, alternative soft ionization techniques, such
as electrospray ionization (ESI) \cite{yamashita:esi,fenn:esi:review} and
matrix-assisted laser desorption/ionization (MALDI)
\cite{karas:maldi,tanaka:maldi,hillenkamp:maldi:review}, are readily available
and would require only small modifications of the presented vacuum chamber.

Lastly, the described experimental setup allows the overlap a $\atom{Ca}$
magneto-optical trap (MOT) with $\atom{Ba}[+]$, $\atom{Ba}\atom{Cl}[+]$, and
$\atom{Yb}[+]$ ions in the RF trap.
The $\atom{Ca}$ MOT leads to a production of undesired $\atom{Ca}[+]$ and
$\atom{Ca}[+]_2$ ions \cite{sullivan:reaction:ca2}, which have an only slightly
lower mass than $\atom{Ba}[+]$.
The operation of the RF trap in a regime that is unstable for
$\atom{Ca}[+]$/$\atom{Ca}[+]_2$ requires a high Mathieu-$q$ parameter
\cite{wineland:bible} for all (desired and undesired) ions.
Thus, it dictates the comparatively low drive
frequency $\Omega_<$ and comparatively high RF amplitude $V_0$.
A high Mathieu-$q$ parameter in turn leads to increased heating rates and
problems for large ion Coulomb crystals.
To tackle this problem, we will explore operation of the RF
trap with different RF drive frequencies being used simultaneously to engineer
unstable regions in the stability diagram for certain masses.
The electronics described in this manuscript already allow such a dual-frequency
mode with the two drive frequencies $\Omega_<$ and $\Omega_>$ and, with it,
we have demonstrated production of ion Coulomb crystals.
Additionally, such a dual-frequency mode could also be advantageous for more exotic
trapping scenarios of ions of largely different masses \cite{Trypogeorgos2014}.
For other applications, operation at a lower $q$ parameter should lead to
improved trapping conditions and performance.

\newcommand{\mysectionacknowledgement}{
\section*{Acknowledgement}
We thank Alexander Dunning and Prateek Puri for critically reading the
manuscript.
This work was supported by the ARO Grant No.~W911NF-15-1-0121, ARO MURI
Grant No.~W911NF-14-1-0378, and NSF Grant No.~PHY-1205311.
}

\newcommand{\mysectionbibliography}{
  \par
  \bibliography{quantumoptics,numerics,publications,misc}
  \bibliographystyle{\mybibstyle}
}

\iftwocolumn{
  \mysectionacknowledgement
  \mysectionbibliography
  \therevisionchanges
}{
  \clearpage
  \mysectionbibliography
  \clearpage
  \mysectionacknowledgement
  \clearpage

  \myfigureschematic
  \clearpage
  \myfigurevoltages
  \clearpage
  \myfiguredrive
  \clearpage
  \myfiguredrivephoto
  \clearpage
  \myfiguresimulationschem
  \clearpage
  \myfiguresimulation
  \clearpage
  \myfigurecontrol
  \clearpage
  \myfigurecontrolphoto
  \clearpage
  \myfigurefeedthrough
  \clearpage
  \myfigureresultsyb
  \clearpage

  \therevisionchanges
}

\end{document}